\documentclass[preprint,preprintnumbers,amsmath,amssymb]{revtex4}

\usepackage{graphicx}% Include figure files
\usepackage{bm}% bold math

\begin{document}

\title{Supplementary Information for:\\
'Spontaneous Skyrmion Ground States in Magnetic Metals'}

\author{U.K. R\"o\ss ler}
\affiliation{
IFW Dresden, P.O. Box 270116, D-01171 Dresden, Germany
}
\author{A.N.\ Bogdanov}
\altaffiliation{Permanent address: 
Donetsk Institute for Physics and Technology,
340114 Donetsk, Ukraine}
%
% \affiliation{
% Physikalisches Institut, Universit{\"a}t Karlsruhe,\\ 
% Wolfgang-Gaede-Str.1, 
% D-76128 Karlsruhe, Germany}
%
%\email{bogdanov@kinetic.ac.donetsk.ua}
%
\affiliation{
IFW Dresden, P.O. Box 270116, D-01171 Dresden, Germany
}
\author{C. Pfleiderer}
\affiliation{
Physik Department E21, 
Technische Universit{\"a}t M{\"u}nchen,\\ 
James-Franck-Strasse, 
D-85748 Garching, Germany}
%
% \affiliation{
% Physikalisches Institut, Universit{\"a}t Karlsruhe,\\ 
% Wolfgang-Gaede-Str.1, 
% D-76128 Karlsruhe, Germany}
%

\date{\today}

\begin{abstract}
Supplementary information for our manuscript, 
entitled 'Spontaneous Skyrmion Ground States of Magnetic Metals', cond-mat/0603103, is presented. 
The physical nature of the gradient terms of 
our generalized micromagnetic model 
for ferromagnets with softened longitudinal fluctuations is explained.
The relationship of our micromagnetic model with the spin fluctuation 
theory of itinerant-electron magnets is discussed.
Experimental estimates of the parameter $\eta$, which accounts 
for an effective reduced longitudinal stiffness, are presented for real materials
from published polarized neutron scattering experiments on EuS, Ni and MnSi.
The available experimental data clearly show that $\eta$ is significantly 
reduced for the latter two systems. 
It is suggested that particle-hole excitations
are at the root of this longitudinal softness in itinerant-electron ferromagnets.
The current status of the experimental evidence supporting 
spontaneous, amorphous skyrmion textures in MnSi and 
other materials is reviewed.
Finally, we also address the general potential of 
skyrmion textures in chiral magnets for other fields of physics. 

\end{abstract}

\maketitle

% \clearpage
% Abstract - extended introduction

\newpage

\section{outline}

An important ingredient for the prediction of spontaneous skyrmion ground states in magnetic metals 
with chiral interactions in our manuscript cond-mat/0603103 \cite{manu06} 
is the assumption that the longitudinal magnetic stiffness is particularly soft.
In section \ref{theo-unde} we consider the theoretical underpinning of this assumption.
The physical nature of the gradient terms of our generalized micromagnetic model 
for ferromagnets with softened longitudinal fluctuations is explained.
As part of the discussion we also address the relationship of our generalized micromagnetic model 
and the spin-fluctuation theory of weak itinerant-electron magnets.
In section \ref{long-real} we review the experimental evidence for $\eta<1$ in real materials.
This is followed by a discussion in section \ref{expe-evid} of available evidence 
that supports a spontaneous skyrmion ground state in the itinerant-electron magnet MnSi.
The supplement concludes with a discussion of the broader implications of spontaneous 
skyrmion ground states in magnetic metals in section \ref{broa-impl}.

\section{Theoretical underpinning of the longitudinal stiffness}
\label{theo-unde}

In the following we address from a theoretical point of view the possiblty of reductions of the longitudinal stiffness in magnetic materials and their importance.
For this purpose we consider the ratio $\eta$ of the longitudinal to the transverse magnetic stiffness, where $\eta$ is identical to that in Eqn.~(1) in the main text \cite{manu06}.
In our discussion we focus on the ordered state of isotropic ferromagnets.
However, the arguments can be generalised in principle to all magnetic materials.
Before entering a more technical level we present an intuitive explanation. 

It is instructive to consider $\eta$ 
at first for isotropic 
local moment ferromagnets (LMFM).
The conventional description of LMFM leads to a
continuum theory for the magnetization
distribution with $\eta \equiv 1$, which is derived 
as the continuum limit of a Heisenberg model 
(for an elementary derivation see, e.g., \cite{akhi}).
As shown by Patashinskii and Pokrovskii \cite{pata74}, 
magnon-magnon scattering in the ordered state 
forces the longitudinal stiffness 
to vanish for vanishing wavevector $q$.
This effect originates alone in the broken continuous 
symmetry of the isotropic Heisenberg magnet, where
$\eta=1$ for $q\to0$ (see also, Zwerger \cite{zwer04}).
On more general grounds we note that
the conservation of total spin,
based on the SU(2) symmetry of the spin system, requires
that a reduction of magnetisation 
amplitude has to occur as a higher order process 
that involves at least two transverse processes.
The finite lifetime of transverse excitations 
in turn results in a reduced longitudinal stiffness.
If there are microscopic processes
limiting the lifetime of transverse excitations in addition to 
magnon-magnon scattering, the longitudinal stiffness 
should be reduced further.
Examples for such processes include: 
(i) interactions of magnons with either intraband 
or interband particle-hole excitations in itinerant magnets,
(ii) interactions of magnons with defects, and
(iii) the interaction of magnons with dipolar components of the magnetisation. 

The stiffness of a magnetic state is a consequence of 
the correlations introduced by these various microscopic processes.
In an ordered or field-polarized state, 
the correlation lengths transverse and longitudinal to the
local polarization direction, $\xi_{\perp}$ and $\xi_{\parallel}$, 
respectively, are generally different.
These correlation lengths
are a direct measure of the stiffness.
Because the decay of longitudinal excitations requires 
higher-order transverse processes, we expect that $\xi_{\parallel}$ 
is always shorter than $\xi_{\perp}$.
Thus,  the longitudinal stiffness is expected 
to be {\em lower} than the transverse stiffness, i.e., 
in real materials we expect $\eta<1$.
This leads us to conclude that 
the longitudinal stiffness in real materials is 
always reduced below pure magnon-magnon scattering ($\eta=1$), 
whence $\eta<1$.
This conclusion is strongly supported by the experimental data 
reviewed in section \ref{long-real},  which gives 
in the local moment ferromagnet EuS ($\eta_{\rm EuS}=0.925$), 
and the itinerant-electron ferromagnets Ni ($\eta_{\rm Ni}=0.65$) and MnSi ($\eta_{\rm MnSi}=0.4$).

We now enter the more technical part of the discussion.
It is helpful to clarify at first the terminology we use.
When we refer to 'micromagnetic models' we refer to a language 
that is in the tradition of the continuum theory of magnetism 
founded mainly by Landau's school in the 1940s.
Micromagnetic models are used in the field of technical magnetism 
and magnetic domains for length scales down to the nanometer scale.
Traditionally micromagnetic models consider only 
changes of the magnetization direction, 
while the amplitude is fixed by the saturation magnetization.
Our generalised micromagnetic model for 
the magnetic stiffness is hence outside the tradition of micromagnetism.
The field-theory underlying 
conventional micromagnetic models is 
the non-linear $\sigma$ model (NL$\sigma$M).
In contrast, the description developed for spin fluctuations 
in itinerant-electron magnets is referred 
to as spin fluctuation theory \cite{lonz85,mori85}.
Spin fluctuation theory is based 
on Fermi liquid theory and the spontaneous 
spin polarisation of the metallic state.
Traditionally, it considers materials where the magnetization 
is free to change direction \textit{and amplitude}.
Both, micromagnetic models and spin fluctuation 
theory can be represented 
in terms of a Ginzburg-Landau functional of 
a 3-vector model for the magnetization distribution.

The task of this discussion is to bring out 
how to modify micromagnetic models 
for a description of itinerant-electron magnets.
The discussion this way also aims 
to clarify the nature of the gradient term in $\eta$.
In the following we first consider 
the relationship of the conventional action of spin fluctuation 
theory with the NL$\sigma$M as the low temperature field-theoretical approximation that corresponds to micromagnetic models. 
This allows to illustrate how to change micromagnetic models 
to become appropriate for itinerant-electron magnets.
We then provide arguments that mechanisms 
beyond magnon-magnon processes, which additionally reduce
the longitudinal stiffness, indeed lead to a significantly reduced 
ratio $\eta<1$ in real materials.

We begin with an outline how to derive 
a corresponding micromagnetic model from 
spin-fluctuation theories for weak itinerant-electron ferromagnets.
The appropriate Ginzburg-Landau functional 
in spin fluctuation theory is developed 
in a classical vector $\mathbf{m}(x)$ 
with three-components \cite{lonz85}, where 
the action (or reduced Hamiltonian) \cite{Zinn-Justin 97} is given by
\begin{equation}
S(\mathbf{m})\,=\,
\int\, d^dx\,[ A\,(\sum_{\alpha}\partial_{\alpha}\mathbf{m})^2
             +\,r\,|\mathbf{m}|^2
             +\,b\,|\mathbf{m}|^4 \dots ]\,.
\label{SFAnsatz}
\end {equation}
The parameters $b$ and $r \equiv a(T-T_C)$ are expansion coefficients with 
$T_C$ the Curie temperature.
The parameter $A$ is the phenomenological exchange stiffness parameter.

To establish the relationship with the micromagnetic model
we transform the action into a form reminiscent of the NL${\sigma}$M 
by writing $\mathbf{m}(x)=m(x)\,\mathbf{n}(x)$ with $|\mathbf{n}(x)| \equiv 1$.
Here $m(x)$ describes the amplitude of magnetization, while $\mathbf{n}(x)$
is a unit vector that accounts for changes of the direction.
In Equation~(\ref{SFAnsatz}) we now replace $\mathbf{m}(x)=m(x)\mathbf{n}(x)$.
This generates two gradient terms in the action
\begin{equation}\label{S0}
S(m,\mathbf{n})\,=\,
\int\, d^dx\,[ A\, m^2(x)\,(\sum_{\alpha}\partial_{\alpha}\mathbf{n})^2
+  A\,\sum_{\alpha}(\partial_{\alpha}m(x))^2 
             +r\,m(x)^2
             +b\,m^4 \dots \,].
\end {equation}
Eqn.~(\ref{S0}) reduces to the NL$\sigma$M 
when the amplitude  is kept constant, 
$m(x)\equiv m_0=\rm constant$, i.e.,  
the fluctuations of the modulus $m$ are neglected.
It follows
\begin{equation}
S_0(\mathbf{n})\,=\,
\int\, d^dx\,[ A\,m_0^2\,(\sum_{\alpha}\partial_{\alpha}\mathbf{n})^2],
\label{NLsM}
\end{equation}
where constant terms have been dropped.
This is the low-temperature field-theoretical approximation 
used in micromagnetism.
To go beyond this approximation, we have to include fluctuations in $m(x)$.
Considering that the longitudinal and transverse stiffness may differ
in the ordered state, we use two different parameters for the two gradient
terms in Eqn.~(\ref{S0}).
This is achieved by introducing the parameter $\eta$.
\begin{equation}\label{S2}
S_1(m,\mathbf{n})\,=\,
\int\, d^dx\,[A\,m^2(x)\,(\sum_{\alpha}\partial_{\alpha}\mathbf{n})^2
+ A\,\eta\,\sum_{\alpha}(\partial_{\alpha}m(x))^2
+r\,m(x)^2
+b\,m(x)^4 \dots] \,.
\end {equation}
As shown in the following, $\eta$ measures 
the ratio of longitudinal to transverse stiffness.
To bring out the nature of the gradient terms 
in the context of our micromagnetic Ansatz, 
Eqn.~(\ref{S2}) can be rewritten 
as a functional of the magnetization vector $\mathbf{m}$
and amplitude $m$ 
\begin{equation}\label{S3}
S_1(m,\mathbf{m})\,=\,
\int\, d^dx\,[A (\sum_{\alpha}\partial_{\alpha}\mathbf{m})^2
-\,(1-\eta)\, A\,\sum_{\alpha}(\partial_{\alpha}m(x))^2
+r\,m(x)^2
+b\,m^4 \dots] \,.
\end {equation}
For $\eta<1$, 
the form of the action $S_1(m,\mathbf{m})$ implies 
that fluctuations of the amplitude lead 
to energy gains as compared
to the case $\eta\equiv 1$.
In other words, the action $S_1(m,\mathbf{n})$ (Eqn\,\ref{S2}) 
describes softened longitudinal stiffness for $\eta<1$.

The fixed modulus version of 
the action given in Eqn.~(\ref{NLsM}), which may be derived by 
approaching the low-temperature limit by taking $r\rightarrow -\infty$, 
$b\rightarrow \infty$, and $r/b=$const, has been
analysed earlier in the context of critical phenomena 
and phase transitions \cite{brez76,pelc76}.
In the present context, we are interested in the 
gradient energy within the magnetically ordered
state at an intermediate length, 
which is set by the competition between
microscopic direct exchange and the chiral
Dzyaloshinskii-Moriya exchange.
It has been shown for the long-distance
behaviour that only the usual gradient energy 
describing the rotation of a fixed-length 
magnetisation is relevant 
as used in micromagnetic theory in Eqn.~(\ref{NLsM}) 
\cite{brez76,pelc76,NoteOnPelc76}.
More specifically, in the renormalization group analysis 
one finds that the gradient term in $(1-\eta)$ in
Eqn.~(\ref{S3}) is irrelevant 
for the critical properties of ferromagnets \cite{NoteOnPelc76}, 
i.e., in the limit of diverging correlation length at $T_C$.
The analysis moreover stresses that only
the transverse fluctuations are relevant to drive 
the phase-transition from the ordered state
into the isotropic paramagnetic state. 
At the same time it is found that 
the irrelevant term with factor $(1-\eta)$ 
has a different scaling behaviour as compared 
to the transverse stiffness.

These renormalization group results 
\cite{brez76,pelc76} assert 
that the generalised gradient energy in our model 
does not change the expected universal 
behaviour of an isotropic ferromagnet 
with collinear ordering.
This is the formal proof that we may 
freely use a generalized version of the gradient energy 
with $\eta < 1$.
As the two gradient terms have differing 
scaling behaviour, a coarse-grained 
effective action after a suitable renormalization 
procedure will always lead to gradient 
terms with $\eta \neq 1$ for the ratio 
of effective longitudinal and transverse stiffness.
Therefore, this ratio will also depend 
on the scale considered under coarse-graining. 
When taken together, these theoretical results 
justify the asssumption that we can 
represent the two different longitudinal and transverse 
correlations lengths {\em in the ordered state}
by the generalization of the gradient energy,
which is embodied in the two parameters $A$ and $\eta$.
For arbitrary values of $\eta > 0$
the gross qualitative properties of
the paramagnetic to ferromagnetic transition 
and the ordered state are unchanged.
However, the term arising for $\eta<1$ 
in the phenomenological theory given by Eqn.~(\ref{S2})
is the most important and simplest term to account 
for effects due to two different lengths 
for the stiffness of the ordered state.
This term changes only non-universal properties.
Essentially, it leads only 
to a shifted ordering temperature \cite{brez76}.
Therefore, the effects due to a softened 
longitudinal stiffness for $\eta < 1$ 
are difficult to trace in the properties of 
a conventional collinear ferromagnet.
However, the generalization for $\eta \le 1$ 
becomes important at intermediate lengths 
in the ordered state when competing magnetic 
couplings are present.
In the case of chiral magnets 
considered in Ref.~\cite{manu06} 
a twisting length arises due to the competition between
Dzyaloshinskii-Moriya couplings and direct exchange.
This gives an additional length scale.
The chiral couplings have 
a tendency to prompt the formation of 
inhomogeneous multiply twisted states.
This is associated with the shift of the ordering 
temperature, according to the 
temperature scale $T_D$ as defined 
in Eqn.~(3) of Ref.~[\onlinecite{manu06}].

For the resulting multiply twisted 
non-collinear magnetization distribution, 
the difference of longitudinal and transverse stiffness 
becomes crucial.
The formation of inhomogeneous 
non-collinear magnetization structures
is faciliated by local shifts of 
the transition temperature that are 
enabled through a softened longitudinal stiffness.
This is the reason why the generalized gradient 
term of our model with $\eta<1$ has to be taken into account. 
We note that the term we consider 
is the leading-order correction needed to generalize
the gradient term in a continuum model for the magnetization.
Thus the conclusions of our study are intuitive and 
of the greatest generality possible.

While our work was going on, 
it has been suggested that higher order 
gradient terms may also 
stabilise multiply twisted structures \cite{tewa06,binz06}. 
However, the possible existence of these terms is specific to
certain materials and difficult to justify microscopically, 
so that the conclusions are not physically transparent and 
lack general relevance. 

Having justified the gradient term in $\eta$ purely from 
a technical point of view,
we now consider how this term evolves 
under coarse graining in a general context.
This serves to certify that the mechanisms listed above, 
which in principle may reduce the longitudinal stiffness, indeed 
will generate an effective model with $\eta<1$.
Following convention, a generalised Ginzburg-Landau 
theory may be written in terms of transverse 
and longitudinal fluctuating fields, $\phi_{\perp}(x)$ and
$m(x)$, respectively  \cite{Nelson 76}.
The longitudinal component or modulus may be further decomposed 
as $m(x)=M+m'(x)$ with a constant expectation value of the 
magnetization $M$ and fluctuations of amplitude  $m'(x)$.
This leads to the action
\begin{eqnarray}
S_e(m',{\phi_{\perp}})& \,=\, & 
\int\, d^dx\,[A\, (\sum_{\alpha}\partial_{\alpha}\phi_{\perp})^2
+ A\,\eta\sum_{\alpha}(\partial_{\alpha}m'(x))^2 \nonumber \\
& & +\,r\,M^2
+\,b\,M^4 \dots \, \nonumber \\
& & +\,r_{||}\,m'(x)^2
+\,r_{\perp}\,|\phi_{\perp}(x)|^2\ \nonumber \\
& & +w_1\,m'(x)|\phi_{\perp}(x)|^2
+w_2m'(x)^3+
+\,b\,(m'^2(x)+|\phi_{\perp}(x)|^2)^4 \dots \,
\label{FullAction}
\end {eqnarray}
with constants $r_{||}=r+32\,b\,M^2$, $r_{\perp}=r+8\,b\,M^2$
$w_1=w_2=16\,b\,M^2$.
We note that the action given in Eqn.~(\ref{FullAction}) is formally the 
analogue to that considered for itinerant-electron magnets at the border of ferromagnetic order \cite{lonz85}.
The limit of a self-consistent coupling of finite $q$ modes to the uniform magnetisation ($q=0$) of 
this model has been studied extensively by Moriya \cite{mori85}, Lonzarich \cite{lonz85} and others.
An important assumption of their analysis is that the response of the system is 
that of a Fermi liquid in the random phase approximation (RPA).

To establish the relationship of our micromagnetic model, 
given by Eqn.~(1) 
in the manuscript [\onlinecite{manu06}]
with Eqn.~(\ref{FullAction}) 
it is in principle necessary to carry out a renormalisation of 
the action $S_e(m',{\phi_{\perp}})$ 
up to the length scales addressed in the micromagnetic model.
Unfortunately, a rigorous renormalization analysis below the
ordering temperature for isotropic $n$-vector models
has not been achieved due to great technical difficulties
in treating mode-mode coupling correctly
\cite{FreySchwabl 94,OConnor 02}.
These difficulties may be traced to the criticality 
of the transverse field components 
along the whole coexistence line below $T_C$ 
and the associated system of Goldstone-modes 
\cite{pata74,Nelson 76,Lawrie 81, Lawrie 85}.
The apparent non-analytic form of the gradient 
energy for $\eta <1$, as seen in Eqn.~(\ref{S2}), 
is likewise expected because of the divergencies related to 
transverse fluctuations
in the ordered state \cite{Lawrie 81, Lawrie 85,OConnor 02}.

While it appears a major challenge to perform 
a rigorous renormalisation 
of isotropic $n$-vector models  
in the ordered state 
that is well beyond the scope of the work reported here, 
it is nevertheless possible 
to track the evolution of $\eta$ 
under renormalisation in a Gedanken calculation.
We begin by noting that 
in a perturbation expansion beyond one-loop-order,
the renormalization of the fields $\phi_{\perp}$ 
and $m'$ scales as powers of 
the correlation lengths $\xi_{\perp}^{-d_{\Phi}}$ 
and $\xi_{\parallel}^{-d_{\Phi}}$, respectively,
where the exponent $d_{\Phi}$ is also known 
as anomalous dimension of the fields
\cite{LeBellac91}.
This implies that renormalizations would, 
in principle, break the isotropy of the model \cite{Nelson 76}.
To guarantee the isotropy of the model under renormalisation,
the longitudinal and transverse field components should instead be 
rescaled consistently by the same factor $\xi_{\perp}^{d_{\Phi}}$.
Thus the action has to be re-paramaterised 
under renormalisation, which is achieved by forcing 
the term in $\eta$ to receive a renormalization by 
an additional factor $(\xi_{||}/\xi_{\perp})^{d_{\Phi}}$.
We note, that along or near 
the co-existence line, the anomalous dimension $d_{\Phi}$ 
is to be considered as an effective exponent of a cross-over region
\cite{OConnor 02}.
Therefore, it will depend on distance 
to the critical points at temperatures $T=T_C$ and 
at $T=0$.
Because the longitudinal correlation length on general grounds
is shorter than the transverse correlation length, 
as discussed above,
$\eta$ must remain less than one 
also after integrating out degrees of freedom 
for the construction of effective coarse-grained theories.

A rigorous proof along the lines of our argument is 
difficult, because the transverse correlation length diverges 
in zero magnetic field, $\xi_{\perp}\to\infty$, 
due to the system of transverse Goldstone modes.
To avoid the problems stemming from 
the divergence of the correlation lengths, we may assume 
that the divergencies at long lengths 
are cut-off by inclusion of an arbitrary small field 
in longitudinal direction.
This field serves also to 
define the longitudinal and transverse directions.
Then, the scaling properties under coarse-graining can be 
connected with spin-fluctuation theories of an isotropic 
itinerant electron ferromagnet in the presence of finite polarization, 
which generally also yields different perpendicular 
and longitudinal correlation lengths as discussed in 
Ref.\,\cite{Takahashi 01,Takahashi 04}.
The form of the spin-fluctuation theory reported in 
Ref.\,\cite{Takahashi 01,Takahashi 04} takes mode-mode coupling 
into account, and the approach assumes that 
the squared modulus of spin-fluctuations is held constant.
Thus, it can be viewed as an approximation for 
an itinerant-electron ferromagnet with maximal 
longitudinal rigidness, similar to the case
of a Heisenberg ferromagnet with local moments (constant modulus).
From Refs.~[\onlinecite{Takahashi 01,Takahashi 04,NoteonTaka04}]
we have that $(\xi_{||}/\xi_{\perp}) < 1$ 
everywhere below the ordering 
temperature at finite polarization, and
$(\xi_{||}/\xi_{\perp}) = 1$, asymptotically at $T_C$.
The relation  $(\xi_{||}/\xi_{\perp}) < 1$ 
given from the spin-fluctuation theory of 
an itinerant-electron ferromagnet in the ordered state 
is the expected result from our general consideration 
on transverse and longitudinal correlation lengths.
Consequently, following our scaling argument, 
$\eta <1$  must hold below the transition temperature 
as a leading order (irrelevant) correction 
in a coarse-grained theory 
for the itinerant-electron ferromagnets.
We note, that this result does not depend on the detailed electronic
structure, say, whether a single-band or multiband system
is considered.

In summary we conclude that the assumption $\eta < 1$ 
should hold quite generally in effective micromagnetic theories,
for the ordered magnetic state whenever amplitude fluctuations 
of the magnetisation are important. 
For fixed moment magnets that are well described 
on a microscopic scale
by a Heisenberg model, the effect may be rather weak 
as it involves higher-order perturbation contributions. 
If the underlying microscopic processes
in a material allow for longitudinal fluctuations 
on short lengths scales, an effective term $\eta < 1$ 
is always expected.

\section{The Longitudinal stiffness in real materials}
\label{long-real}

In the following we review the transverse and longitudinal magnetic stiffness 
observed experimentally by polarised neutron scattering at the border of magnetic order in real materials. 
In the context of our micromagnetic model we are interested, 
if the longitudinal stiffness is reduced on \textit{intermediate} lengthscales. 
For this purpose we consider the ratio $\eta$ of the longitudinal 
to the transverse magnetic stiffness, where we refer 
to section \ref{theo-unde} for a more detailed theoretical underpinning of the analysis.
As materials we consider the cubic systems EuS, Nickel and MnSi, 
which exemplify local-moment ferromagnetism, itinerant-electron ferromagnetism 
and weak itinerant-electron ferromagnetism, respectively. 
In MnSi we the presence of weak chiral interactions do not alter the outcome of these estimates.

In the three compounds considered here 
the wavevector dependent susceptibility 
follows the Ornstein-Zernicke form $\chi_i^{-1}(q)\approx c_i(\kappa_i^2+ q^2)+\ldots$ over major portions of the 
Brillouin zone, where the index $i$ denotes transverse ($i=\perp$)
and longitudinal components ($i=||$), respectively.
This form is consistent with the generalised spin fluctuation model given in Eqn.\,(\ref{FullAction}).
The parameter $\kappa_i=1/\xi_i$ is the inverse correlation length and the parameter $c_i$ the stiffness.
The stiffness ratio $\delta=c_{||}/c_{\perp}$ permits 
a comparison of different materials without need for complicated normalisation procedures of the data.
For the cubic systems considered here 
the total susceptbility near $T_C$ is expected 
to be consistent with a 3-component vector model \cite{alsn76}, where
$\chi_{tot}=2\chi_{\perp}+\chi_{||}$ accounts for 2 transverse and 1 longitudinal mode.
For a total susceptibility of the Ornstein-Zernicke form 
it follows that $\chi_{tot}^{-1}=c_{tot}(\kappa^2_{tot}+q^2$), 
where transverse and longitudinal contributions are given by
$\chi^{-1}_{\perp}=c_{\perp}(\kappa^2_{\perp}+q^2) = (3/2) \chi_{tot}$
and 
$\chi^{-1}_{||}=c_{||}(\kappa^2_{||}+q^2) = 3 \chi_{tot}$, 
respectively.
Thus, when the longitudinal and transverse stiffness 
are the same it follows $\delta=2$.
The parameter $\eta$ introduced above is given by $\eta=\delta/2$, i.e.,
$\eta$ is the stiffness ratio per mode.

By definition  $\chi(q)$ is equivalent 
to the frequency integrated scattering intensity.
To obtain the stiffness ratio of EuS we have taken the slopes of the inverse integrated scattering intensities, and thus $\chi_i^{-1}(q)$, shown in Fig.\,6 in Ref.\cite{boen02}.
Here we obtain $\delta_{\rm EuS}=c_{||}/c_{\perp}\approx1.85$.
This corresponds to $\eta_{\rm EuS}\approx0.925$.
Taking likewise the ratio of the slopes of the integrated scattering intensity for the itinerant-electron ferromagnet Nickel shown in Fig.\,10 in Ref.\cite{boen91} we obtain
$\delta_{\rm Ni}\approx1.3$ or $\eta_{\rm Ni}\approx0.65$
Finally, for MnSi 
the available spin-polarised neutron 
data yield $\delta_{\rm MnSi}\approx0.8$ 
\cite{sema99,boen06}, i.e., $\eta_{\rm MnSi}\approx0.4$.
Hence, in the range where data are available,
the longitudinal stiffness in Ni and MnSi
is reduced by 35\% and 60\%, respectively, 
Because the $q$-range considered in Ni and MnSi is in a regime 
where neutron scattering shows an abundance of 
spin-flip particle-hole excitations, 
we  conclude that particle-hole excitations are dominant 
in causing the reduction of the longitudinal stiffness of Ni and MnSi.

The conjecture that particle-hole excitations are dominant 
in reducing $\eta$ is supported further by the characteristic frequency dependence, 
which reflects on the mechanisms that determine the transverse and longitudinal stiffness.
For local moment systems such as EuS, 
it is given by the R{\'e}sibois-Piettes function.
In contrast for itinerant-electron magnets, an additional softening 
has been noticed (see, e.g., Fig.~13 in Ref.~\cite{boen91} for Ni).
A simple measure of the frequency dependence and thus $\eta$ may 
be the spin fluctuation temperature, $T_{sf}$, defined from 
the width of the spin fluctuation relaxation frequency spectrum.
The ratio of Curie temperature to spin fluctuation temperature, 
$\tau=T_C/T_{sf}$, may be used as 
an estimate for the magnitude of $\eta$,
notably, the smaller the ratio $\tau$  the smalller $\eta$.
This suggestion is consistent with the 
experimentally observed reduced longitudinal stiffness 
in Ni and MnSi, discussed above.

When taken together, the experimental evidence establishes 
beyond doubt that many materials exist, 
which may be described by the regime $\eta<1$ of our model.
Moreover, the experimental data identifies MnSi, which supports 
chiral interactions, as a prime candidate for 
a spontaneous skyrmion ground state, as discussed in section \ref{expe-evid}.
Interestingly the experimental data even shows $\eta<1$ in the local moment magnet EuS.
This is not a suprise, since a multitude of mechanism 
exists that reduce the longitudinal stiffness.
As a consequence, spontaneous skyrmion phases may exist in chiral local moment magnets, albeit they may be much less stable.

\section{Possible evidence of sykrmion ground states in MnSi}
\label{expe-evid}

As the main result of our study we predict the spontaneous formation of a skyrmion ground state in a temperature interval prior to helical order in all materials with soft magnetisation amplitude and weak chiral interactions.
Bulk compounds which develop long-wavelength helical order of the kind considered here are, for instance, MnSi \cite{mnsi-DM}, FeGe \cite{lebech89} and (Fe,Co)Si \cite{beil83}, where the modulation wavelengths are $\lambda_{\rm MnSi}\approx180${\AA}, $\lambda_{\rm FeGe}\approx700${\AA} and $\lambda_{\rm (Fe,Co)Si}\approx 300${\AA}, respectively.
In these materials the helical modulation is driven by Dzyaloshinsky-Moriya chiral interactions that originate in the lack of inversion symmetry in their cubic B20 crystal structure. 
Being transition metal compounds the magnetic order 
is due to itinerant electrons, although this has only been shown experimentally in MnSi \cite{tail86}.
As discussed above polarised neutron scattering data exist for MnSi  \cite{sema99} that establish a reduced longitudinal stiffness ($\eta_{\rm MnSi}=0.4$).
Thus at least MnSi meets all the requirements for the formation of the predicted spontaneous skyrmion ground states.

The helical order in MnSi, FeGe and (Fe,Co)Si may be seen 
most easily in small-angle neutron scattering (SANS) experiments, 
where sharp magnetic satellites were observed.
In a small temperature interval above the onset of helical order,  
SANS data for MnSi and FeGe show 
neutron scattering intensity  located 
on the surface of a sphere in reciprocal space
as an additional feature\cite{lebech89,lebe93}.

The radius of this sphere corresponds 
to the same modulation length as the helical order.
The sphere of scattering intensity has therefore traditionally been interpreted as fluctuating helical order.
However, in these studies it was not considered that the scattering intensity may also be related to a frozen-in form of order.
In this respect it is interesting to note, that the scattering intensity on the surface of the sphere is perfectly consistent with an amorphous  array of randomly oriented cylindrical skyrmion tubes.

To better identify of the ring of scattering intensity 
we present a rough estimate of the temperature interval $\Delta T=T_D-T_C$,
for which the skyrmion ground state is expected in MnSi.
The lattice constant of MnSi is $a_{\textrm{MnSi}}=4.55$\,{\AA}. 
There are four formula units per unit cell, so that 
the number-density is $N/V = n = 4.246\cdot10^{28} {\rm m}^{-3}$.
The effective moment determined from the Curie-Weiss temperature dependence of the susceptibility is $p_{\textrm{eff}} = 2.2 \mu_{B}$  \cite{Yasuoka 78}
and the susceptibility $\chi = C_{\textrm{CW}}/(T-T_C)$ where
$C_{\textrm{CW}} = n \mu_{0} \mu_{B}^2 (p_{\textrm{eff}}^2) / (3 k_B) = 0.534$~K.
The upper limit of the skyrmion transition temperature is  
$T_h=T_C + (1/2) (D/(2A)) (D/a)$ 
as given in Eq.(2) in Ref.~[\onlinecite{manu06}].
These parameters have to be taken 
from experimental data, notably Ref.~[\onlinecite{Grigoriev 05}].
The parameter $a$ is given by the initial
susceptibility and may be estimated from the Curie-Weiss susceptibility:
$a=1/(2C_{\textrm{CW}})  = 0.936\rm\,K^{-1}$.
The modulus of the helical propagation vector
is $q_0=(D/(2\,A))=0.039$~\AA$^{-1}$.
In the following we use temperature as unit of energy-density.
The exchange stiffness $A$ may then be estimated from 
$A \simeq (a_{\textrm{MnSi}}^2)\,T_C = 
50\cdot 10^{-3}$~eV \AA$^{2} =$ 587 K \AA$^{2}$.
The strength of the DM-coupling $D$ may be estimated 
from $q_0$ as $D = 1.9\cdot 10^{-3}$~eV~\AA $= 22$~K~\AA.
These parameters finally result in
$T_h = T_C + (1/2)  q_0  (D/a) = T_C + q_0  D  C_{\textrm{CW}}$
% $(29.5 + 0.039 *  22 *  0.534)$~K 
$=(29.5 + 0.46)$~K.
The ordering temperature of the skyrmion phase 
(according to Eq.(3) in Ref.~[\onlinecite{manu06}]) 
is obtained by doubling the difference between $T_C$ and $T_h$ so that
$\Delta T = T_D-T_C = 0.9$K.
This value of $\Delta T$ 
gives an order-of-magnitude estimate for the temperature 
interval of the stable skyrmion phase in MnSi as shown
in the phase diagram of Fig.~2 in our manuscript \cite{manu06}.

Detailed studies of the specific heat (Fig. 6 in Ref.\cite{pfle01b,pfle97}) and thermal expansion \cite{mats82} of the paramagnetic to helical transition in MnSi show the presence of a spike at $T_h$ and a broad shoulder in a temperature interval of order $\sim$1\,K above this transition.
Detailed SANS studies of the ring of intensity 
as function of temperature suggest 
that the ring extends over the same temperature 
interval where the shoulder in the specific heat is seen \cite{lama06}.
Interestingly the temperature interval corresponds 
fairly well with that predicted for the skyrmion phase.
In combination, the broad shoulder in the specific heat clearly signals the presence of quasi-static order already above $T_h$ that may be either related to randomly oriented helical domains or an amorphous texture of skyrmions.
An important additional aspect
that has been emphasized in \cite{pfle01b} is, 
that the shape of the specific heat anomaly 
in very high purity samples is quite different, yet, 
still supports the possible existence of a precursor phase.
However the shoulder is observed for samples with largely varying residual resistivity ratio \cite{pfle97}.
It is possible that the dependence on extreme sample purity 
signals that the stabilisation of the quasistatic 
order sensitively depends on residual defects in the material.

The spontaneous formation of skyrmion ground 
states allows to explain many of the mysterious 
properties of MnSi at high pressure and low temperatures.
It was recently reported by one of us (CP and collaborators), 
that the metallic state of MnSi changes as function of pressure from a quadratic temperature  dependence of a weakly spin-polarized 
Fermi liquid to a $T^{3/2}$ resisitivity that is stable over a remarkably large pressure range \cite{pfle01,pfle03,doyr03,pedr05}.
Three aspects appear to be inconsistent with the present day understanding of metallic magnetism:
First, as pointed out in Ref.\,\cite{pfle01} 
a $T^{3/2}$ resistivity is normally observed 
in spin glasses and amorphous ferromagnets, 
where it is explained by a diffusive motion of 
the charge carriers \cite{rivi}.
However, for the high-purity MnSi single-crystals investigated  
the behaviour of a moderately enhanced Fermi liquid is expected by all accounts.
It was therefore emphazised that the diffusive charge carrier motion would have to 
be intrinsic.
Second, the $T^{3/2}$ resistivity is extremely stable as function of pressure, where it has been seen up to at least 45\,kbar\,$\approx3p_c$. 
This suggests the formation of an extended new phase in contrast to a cross-over phenomenon normally observed at quantum phase transitions.
Third, the $T^{3/2}$ resistivity is extremely stable as function of magnetic field, where experiment shows that it collapses abruptly only above a certain critical field related to a so-called itinerant-electron metamagnetic transition (MMT) \cite{thes97}.
At the highest pressures the $T^{3/2}$ resistivity extends at least up to 1\,T (cf Fig. 3 in Ref.\,\cite{doyr03}) which substantially exceeds the field needed to collapse helical order $B_c\approx0.6T$. We note that $B_c$ studied by various magnetic probes is insensitive to pressure.

As a possible resolution to these contradictions we suggest that the Fermi liquid to non-Fermi liquid transition in MnSi may signal a transition between helical order and an amorphous skyrmion ground state.
A spontaneous amorphous skyrmion ground state provides a simple \textit{intrinsic} mechanism that explains why the transport properties of an amorphous ferromagnet are seen even in a high purity material in the absence of disorder or frustration.
The leading order effect of pressure in this scenario would be to generate a slight mode softening that reduces the stability of helical order, but not to collapse the magnetic moment on local scales completely as normally assumed in quantum phase transitions.
Because the skyrmions form as a genuine, stable ground state, it would finally not be difficult to explain, a priori, 
the extent of the $T^{3/2}$ resistivity behaviour 
in the temperature-pressure phase-diagram.
Also, it has been shown 
that the core of cylindrical skyrmion tubes is extremely 
stable against external fields \cite{bogd94}.
Thus, if the $T^{3/2}$ resistivity is due to an amorphous skyrmion texture 
it is expected that the behaviour survies to much higher fields.
The signature of such an amorphous skyrmion ground state 
in neutron scattering would be intensity on the surface of a small 
sphere in reciprocal space, similar to that seen 
at ambient pressure above $T_h$.

Recent neutron scattering studies of the magnetism as function of 
pressure indeed  
show that large ordered moments survive 
deep into the phase where the $T^{3/2}$ resistivity is observed \cite{pfle04}.
The ordered moments are organised such that they 
lead to scattering intensity on the surface of a sphere 
in reciprocal space.
There is, however, a very important difference 
with the intensity on a sphere seen at ambient pressure near $T_h$.
The ambient pressure intensity is isotropic, while broad maxima are observed for $\langle110\rangle$ at high pressure. 
The  $\langle110\rangle$ direction is unusual for this cubic magnet.
This is so, because (i) the leading magnetocrystalline anisotropy 
favours easy magnetic axes only in $\langle111\rangle$ 
or  $\langle100\rangle$ directions. And, (ii), in itinerant 
d-electron magnets as MnSi with very weak anisotropy, it 
is very unlikely that higher-order anisotropy 
contributions do play any role.
In fact, the easy axis of the helical order 
in MnSi is  $\langle111\rangle$ and in FeGe  $\langle100\rangle$.
Thus, if there was randomly oriented helical 
order at high pressure in MnSi, it would be very difficult 
to explain the broad maxima along  $\langle110\rangle$.
In contrast, for a condensate of skyrmion tubes 
the odd maxima for $\langle110\rangle$ are easily explained, 
if we suppose that the easy magnetic axis of MnSi, $\langle111\rangle$, 
does \textit{not} change as function of pressure, 
while the magnetism changes from a helical 
modulation along $\langle111\rangle$
to an amorphous texture with cylindrical skyrmion tubes 
having their axes preferentially along $\langle111\rangle$.
For cylindrical skyrmion tubes 
that are stratified along $\langle111\rangle$
neutron scattering intensity in reciprocal space
is expected on great circles perpendicular 
to  $\langle111\rangle$ directions. 
As all $\langle111\rangle$ directions produce intensities 
along such great circles, these will intersect
and produce maxima of intensity in the $\langle110\rangle$ directions.
The available experimental data are consistent with this possibility.

However, in contrast to the remarkable extent of 
the $T^{3/2}$ resistivity under pressure and magnetic field, 
the sphere of scattering intensity under pressure is seen only 
in a fairly small portion of the phase diagram.
A possible explanation discussed in Ref.\,\cite{pfle04} is that 
the textures are dynamically destabilized and become 'invisible' 
in the neutron scattering experiment.
Such a slowly meandering form of magnetic order 
could persist on a time scale that it still 
much slower than the time scale relevant 
to the electrical resistivity.
A liquid instead of a frozen skyrmion phase in this pressure range 
provides a natural explanation for the difference in the 
transport measurements and the visible static magnetic order.
Such a liquid skyrmion phase would not be unusual and may be 
analoguous to liquid vortex phases in superconductors \cite{blat94}.
It will, in any case, require a host of novel 
experimental techniques to settle this issue unambiguously.
For instance, real space imaging of helical 
order in thin (Fe,Co)Si layers was recently 
reported using Lorentz force microscopy (LFM) \cite{uchida06}.
However, samples that can be studied 
with LFM have to be atomically thin 
so that uncontrolled mechanical strains and 
their interplay with the magnetism represents a major concern.
As regards LFM in MnSi we note, moreover, 
that even state-of-the-art LFM does not 
offer sufficient resolution yet 
to image the magnetic structure with sizes
of the twisting length 18~nm.
We believe that, as an important aternative 
to real-space imaging, the most promising method 
in the immediate future will 
be full three-dimensional polarisation analysis 
for neutron scattering. This is presently 
under development under the synonym mu-PAD.
%\textbf{Webpage des Instruments am FRM-II zitieren?}

\section{On the broader implications of spontaneous 
skyrmion ground states due to chiral interactions}
\label{broa-impl}

Skyrmion states in field theories
have attracted great interest because they represent stable, particle-like 
objects with a nontrivial topology that originates in fundamental 
symmetries \cite{makh93,witt83}. 
For instance, Skyrme's field theory for elementary
nuclear particles has been applied to 
condensates of a few skyrmions \cite{braa86} 
and extended lattice structures \cite{kuts84} 
as models of ordinary nuclei and compact stars, respectively.
Our prediction of spontaneous skyrmion ground states 
in chiral magnets is connected to the
fundamental questions about the relation between 
continuum theories, 
the countable nature of particles and the possibility
to relate both aspects by topology in extended matter systems.
In this more general context  
the role of chirality and chiral interactions, which are possible
in condensed matter systems \cite{Let95}, 
has not yet been appreciated.
Chiral interactions in condensed matter systems offer 
an exceptionally rich setting, because they exist 
in many different contexts, e.g., 
(i) spin-orbit interactions in non-centrosymmetric magnetic materials,
also referred 
to as Dzyaloshinsky-Moriya (DM) interactions \cite{Dz64,izyu84},
(ii) in non-centrosymmetric ferroelectrics \cite{sosn82,hala98},
(iii) for certain structural phase transitions \cite{lifs41,tole87},
(iv) in chiral liquid crystals \cite{wrig89}, or 
(v) in the form of Chern-Simons terms 
in gauge field theories \cite{murt03,chen93}.

The example of an extended magnetic skyrmion texture 
shown in Figure~4 of Ref.~[\onlinecite{manu06}] 
demonstrates the condensation of 
particle- or string-like localized objects.
There are, however, further intriguing possibilities 
beyond these extended ground-states.
Magnetic skyrmions may also arise as 
non-linear localized {\em excitations}
both in homogeneously ordered  systems \cite{bogd94}
and in the paramagnetic state. 
This possibility has not yet been considered
in theories on the metallic state 
of systems with broken inversion symmetry.
In particular, the description of skyrmion excitations on a paramagnetic
background would be very similar to that of particles
generated in a vacuum state.
While being large and localised, they are subject to an intrinsic 
length scale related to the twisting length of the system.
This lengthscale implies non-dispersive behaviour
unlike propagating waves.
Moreover, skyrmions in condensed matter systems may allow
to study processes like particle generation and annihilation, 
scattering between particles, or particle confinement.

Fundamental processes of this kind are usually the subject of 
field-theories in particle physics only.
Images of such processes for
elementary particles can therefore be inferred 
at best only indirectly.
In chiral magnetic condensed matter, 
these fundamental processes 
take place in a continuum distribution of magnetization 
that may be visualized in a solid state laboratory 
by modern magnetic real- and reciprocal space imaging methods.
In fact, because the spontaneous skyrmion ground states 
predicted here represent large (several hundreds \AA),
but localized non-linear spin-fluctuations, 
they would provide the first cases 
of skyrmions in a condensed matter system, 
where detailed imaging of such processes may be possible.
With this perspective in mind, it is gratifying
that novel chiral crystals have recently
been found, where superconductivity coexists
with magnetic order, as in
the novel heavy-fermion superconductors CePt$_3$Si 
\cite{Bauer04} and UIr \cite{Akazawa04}.
In these materials different types 
of long-range order coexist, so that it may be possible to study 
interactions between skyrmionic excitations 
in the associated order-parameter fields 
which in turn have different symmetries.
In fact, chiral condensed matter systems could 
eventually be used to mimick a universe with different 
types of  particles and their interactions.

In spite of their low symmetry, metallic magnets 
with broken inversion symmetry are rather common. 
In particular, Dzyaloshinsky-Moriya interactions
are induced by surfaces 
in {\em all} magnetic nanostructures \cite{Fert90,PRL01}. 
Thus magnetic nanostructures can endow science 
with an arena, where creation, interactions, 
and condensation of particle-like states 
in a continuous field can be engineered for experiment.


\begin{thebibliography}{99}
\bibitem{NoteOnPelc76}
In Refs.~\cite{brez76} the irrelevance of 
``dimension-2 operators'' is remarked. 
This means that there can be two parameters 
in the gradient energy (of dimension-2) of a generalized NL$\sigma$M
\cite{pelc76}.
These are the terms in Eqn.(2) of Ref.~[\onlinecite{pelc76}] 
with parameters  $f \propto A$ and $g_1$, which plays the role of $1-\eta$. 
%
By recursion relations, it is shown that for the ferromagnetic ordering, the 
terms related to $g_1$ are irrelevant (see Eqs.(4) and (5) in \cite{pelc76}).
\end{thebibliography}
\end{document}